# The Tsallis Entropy Barrier or the Roundness Barrier Based Dynamic Stochastic Resonance ---A New Family of SR ?

Xiangjun Feng

World Chinese Forum on Science of General Systems (WCFSGS)*

*Wexford, Pennsylvania, USA*

*WCFSGS@gmail.com*

***Abstract*— The Tsallis entropy barrier or the roundness barrier based dynamic stochastic resonance mechanisms are put forward and simulated. The systems with various Tsallis $q$ values exhibit the effects of emergence as a result of the noise-induced cooperative phenomena.**

***Key Words*—Stochastic Resonance, Tsallis Entropy, Tsallis Entropy Barrier , Roundness, Roundness Barrier, Emergence.**

## I. INTRODUCTION

The stochastic resonance mechanism (SR) was found in 1981 by Professor Roberto Benzi et al [1]. Typical science and technologies of system science such as the uncertainty optimization [2], the emergence as a result of noise -induced cooperative phenomena [3] , the noise-induced order or the self-organization [4], the catastrophe in empirical world or a resonance point in mathematics [5] etc. were implemented or demonstrated in various systems with stochastic resonance mechanisms. It was stated that SR has been implemented by mother nature on almost every scale [6]. I have been deeply involved in the research of the stochastic resonance since 1994 [7]. Recently I mentioned the concept of Stochastic Resonance School of System Science [8][9][10]. In this paper, I try to put forward an idea of greatly expanding the family of dynamic stochastic resonance mechanisms in a natural way. The widely studied paradigmatic dynamic stochastic resonance mechanism is characterized with an over-damped motion of a generalized particle in a bistable double-well potential. Obviously , it is still necessary to explore and unveil more natural or artificial forms of dynamic stochastic resonance in spite of the facts of that non-dynamic stochastic mechanisms with a threshold have been found [11], and improved information-theoretic measures were shown with such a system at the optimal point of the uncertainty or the noise [12]. In a paper [13], I put forward the concept and formula about the roundness of the generalized system which describes the degree of isotropy or uniformity of the probability distribution of a generalized system. In another paper [14], I pointed out that the averaged roundness is actually the special case of Tsallis entropy with $q = 2$. Recently I generalized the formula of the roundness based on the concept of that moments of the probability values themselves are measures of the anisotropy or non-uniformity of a generalized system, and the generalized formula is proportional to Tsallis entropy for any given Tsallis $q$ value [15]. In the next section of this paper, I will show that the roundness distribution or the Tsallis entropy distribution with certain Tsallis $q$ value forms a barrier for the contradictions in motion , and this natural barrier may serve as an essential component of the stochastic resonance mechanisms for generalized systems. In this paper, I will only discuss the cases of that the Tsallis $q$ values are integers not less than 2.

* World Chinese Forum on Science of General Systems (WCFSGS) is an online journal with ISSN: 1936-7260 and also an emerging organization of learning with the web address of http://www.aideas.com The official languages of WCFSGS are both English and Chinese.

## II. THE TSALLIS ENTROPY BARRIER OR THE ROUNDNESS BARRIER BASED STOCHASTIC RESONANCE MECHANISMS

The Tsallis entropy [16] or the roundness is proportional or equals to the $E_n$ expressed as

$$E_n = \frac{1}{q-1}(1 - p_1^q - p_2^q - ... - p_n^q) \quad (1)$$

where $p_i$ is the probability of the $ith$ component of a generalized system, and $i = 1,2,...,n$. The $q$ is the Tsallis $q$ value. In this work, the $E_n$ is re-defined as the Tsallis entropy barrier or the roundness barrier. I am particularly interested in the special case of $n = 2$. For this special case,

$$E_2 = \frac{1}{q-1}(1 - p^q - (1-p)^q) \quad (2)$$

where $p$ is the weight or the probability of one side of a contradiction in motion. The normalized distributions of the Tsallis entropy barrier or the roundness barrier about the weight or probability $p$ for various integer $q$ values are shown in Figure 1.

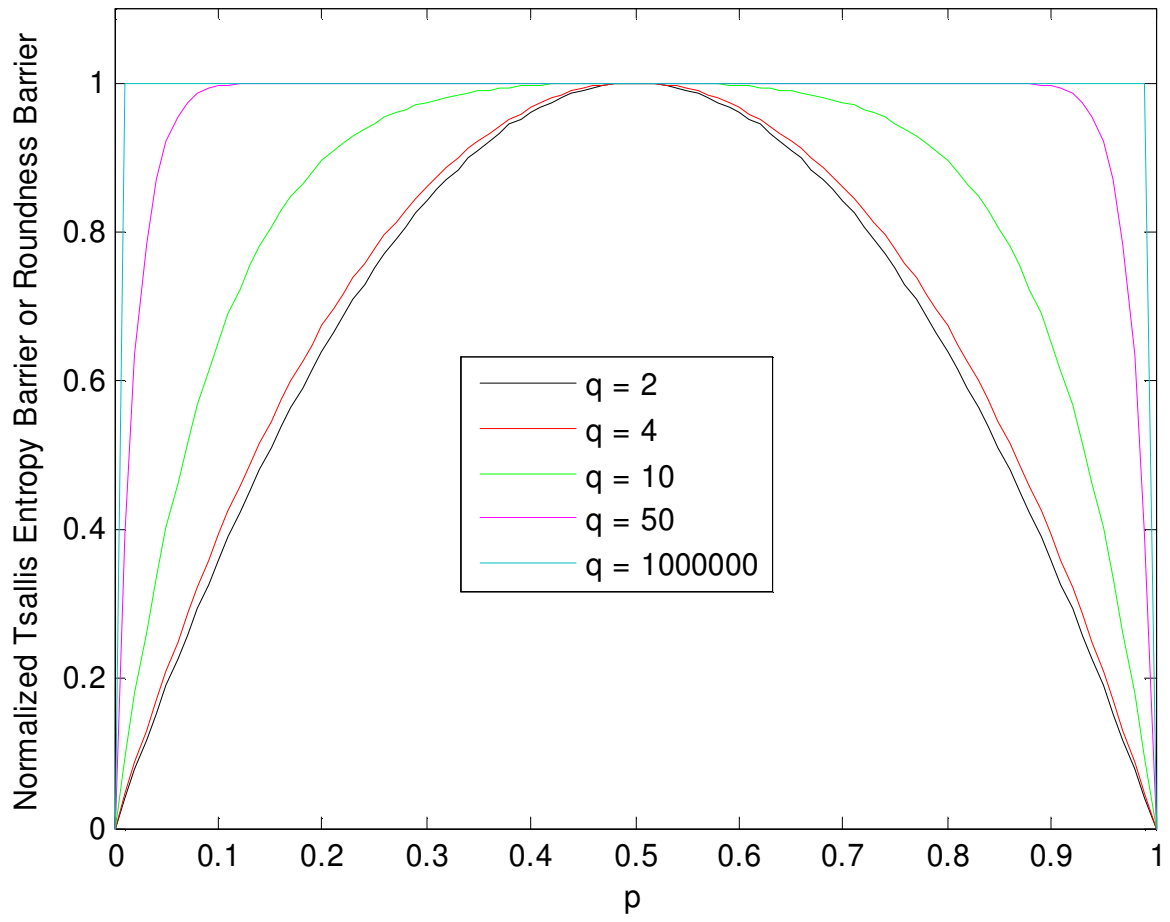

Figure 1. The normalized distributions of the Tsallis entropy barriers or the roundness barriers about the weight or the probability $p$ for various Tsallis integer $q$ values.

For contradictions in motion, both sides take occupying the absolute ruling position ( $p->1$ ) as the target. This target makes various forms of the absolute anisotropy or non-uniformity as the absorbing centers of the contradictions in motion. On the other hand the maximal entropy, maximal roundness, maximal isotropy, and maximal uniformity are the natural tendency under various constraints [13][17]. Therefore the distributions of Tsallis entropy or the roundness become barriers for both sides competing with each other to gain the absolute ruling position. The barrier of the entropy exists everywhere in the motion of a contradiction in which both sides must overcome the natural tendency of maximizing the entropy. I redefined the Tsallis entropy as a barrier not only because the absorbing centers of the contradiction in motion are in the opposite direction of the absorbing center for the natural tendency of maximizing the entropy but also because the distribution of Tsallis entropy is a wonderful description about the barrier shown in a dynamic process. The dynamic process is described as follows. For the side currently at a very weak position ( $p->0$ ) and aiming at the absolute ruling position ( $p->1$ ), It has to experience something like "climbing a mountain" and then "going down the mountain". At the early stage, the resistance for the side to gain any weight from its rival is very great. As its weight keeps increased, the resistance becomes smaller and smaller. When the weight reaches the point of $p=0.5$ where the side becomes equally important as its rival, the side arrives at the peak of the barrier. When the weight is larger than 0.5, instead of suffering from a resistance, it begins to get a force of assistance which helps it to run towards the dreaming point of $p=1$. As the weight increases further more, the force of assistance will also be increased until the side reaches the absolute ruling position of $p=1$. The main characteristics of such a barrier can just be well described by the distribution of $E_2$ which is the Tsallis entropy barrier or the roundness barrier for the case of $n=2$. Keeping in mind the concept of the Tsallis entropy barrier or the roundness barrier, I established the following differentiation equation for a new family of dynamic stochastic resonance mechanisms.

$$\frac{dy}{dt} = \frac{q}{q-1}(y^{(q-1)} - (1-y)^{(q-1)}) + a\xi(t) + b\sin(2\pi ft)$$

and

$$y=0, \text{ if } y<=0, \quad y=1, \text{ if } y>=1. \quad (3)$$

where $y$ is a generalized position of a generalized particle and is not necessarily a weight or a probability anymore , $q$ is an integer corresponding to Tsallis $q$ value, $a\xi(t)$ is the uncertain noise with the sigma $a$, $b\sin(2\pi ft)$ is the deterministic periodic force with $b$ as the amplitude and $f$ as the frequency, and $t$ is the time. In next section, the initial results of the simulation with MATLAB will be presented.

## III. THE INITIAL RESULTS OF THE SIMULATION WITH MATLAB

I made a simulation with MATLAB about the Tsallis entropy barrier or the roundness barrier based stochastic resonance mechanisms. The Tsallis $q$ values used were 2, 4, 10, 50 and 1000000. For each case, the simulation results were summarized and presented with a figure consisting of five subplots . The first subplot described the effects of emergence in the output when both the deterministic periodic force and the optimized uncertainty or the uncertain noise appear in the input and induce a cooperative phenomena. The second subplot was the result of filtering the output presented in the first subplot with a simple threshold- filter. The transfer function of the threshold-filter is as follows.

$$Y=1, \text{ if } X>0.5, \text{ and } Y=0, \text{ if } X<=0.5. \quad (4)$$

where $X$ is the input and $Y$ is the output.

The third subplot was for the total input of the deterministic periodic force plus the uncertain white Gaussian noise. The forth subplot is for the output in the case when there is only the deterministic periodic force as the input, and the fifth subplot was for the output in the case when there is only the uncertain white Gaussian noise as the input. The five figures for Tsallis $q$ values of 2, 4, 10, 50 and 1000000 are shown in Figure 2-6.

It is demonstrated with all the figures that there are effects of emergence in the output as the results of the noise-induced cooperative phenomena.

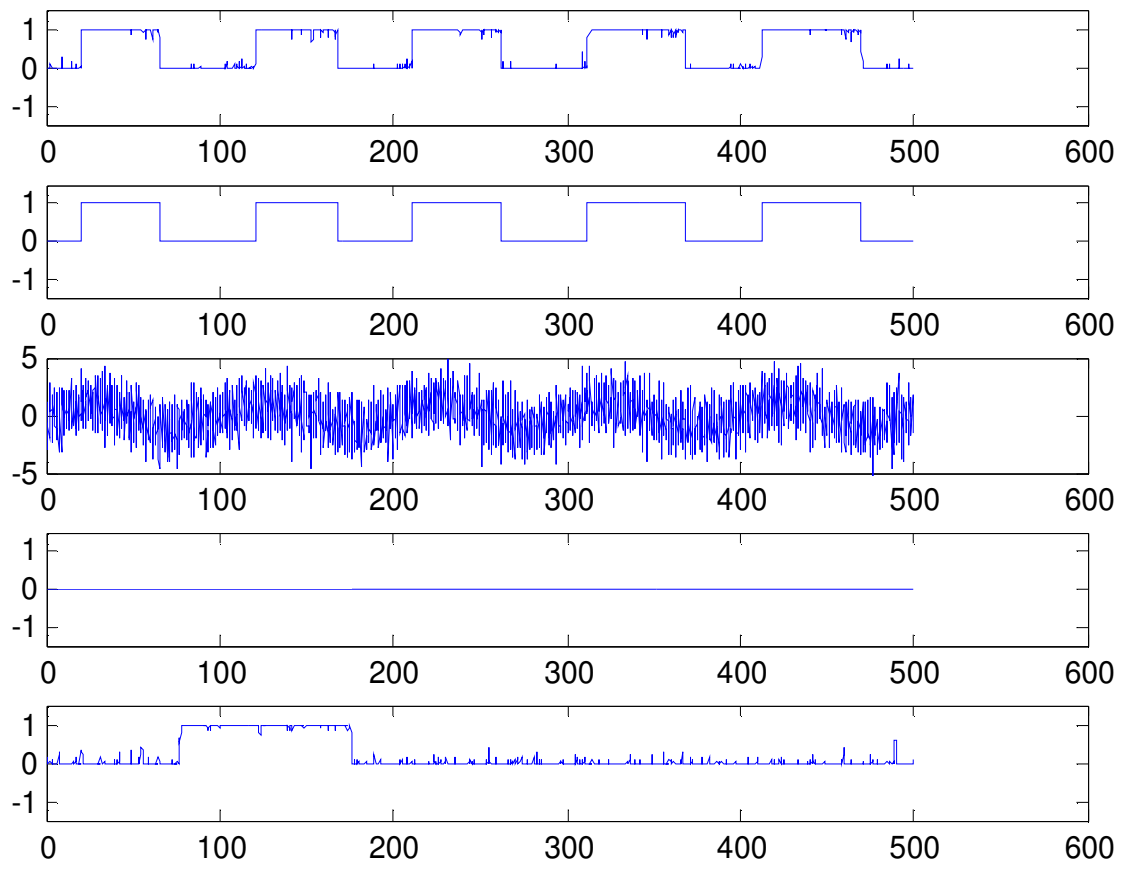

Figure 2. The simulation results for the Tsallis barrier or the roundness barrier based stochastic mechanism with $q = 2$, $f = 0.01$, $a = 1.3$, $b = 1.05$. The subplot 1 is for the output in the case when both the deterministic periodic force and the uncertainty or the uncertain white Gaussian noise appear in the input and induce a cooperative phenomena. The subplot 2 is for the output of the simple threshold-filter taking the output shown in subplot 1 as the input. The subplot 3 shows the total input of both deterministic periodic force and the uncertainty or the uncertain white Gaussian noise. The subplot 4 is for the output in the case when there is only the deterministic periodic force as the input, and the subplot 5 is for the output in the case when there is only the uncertain white Gaussian noise as the input.

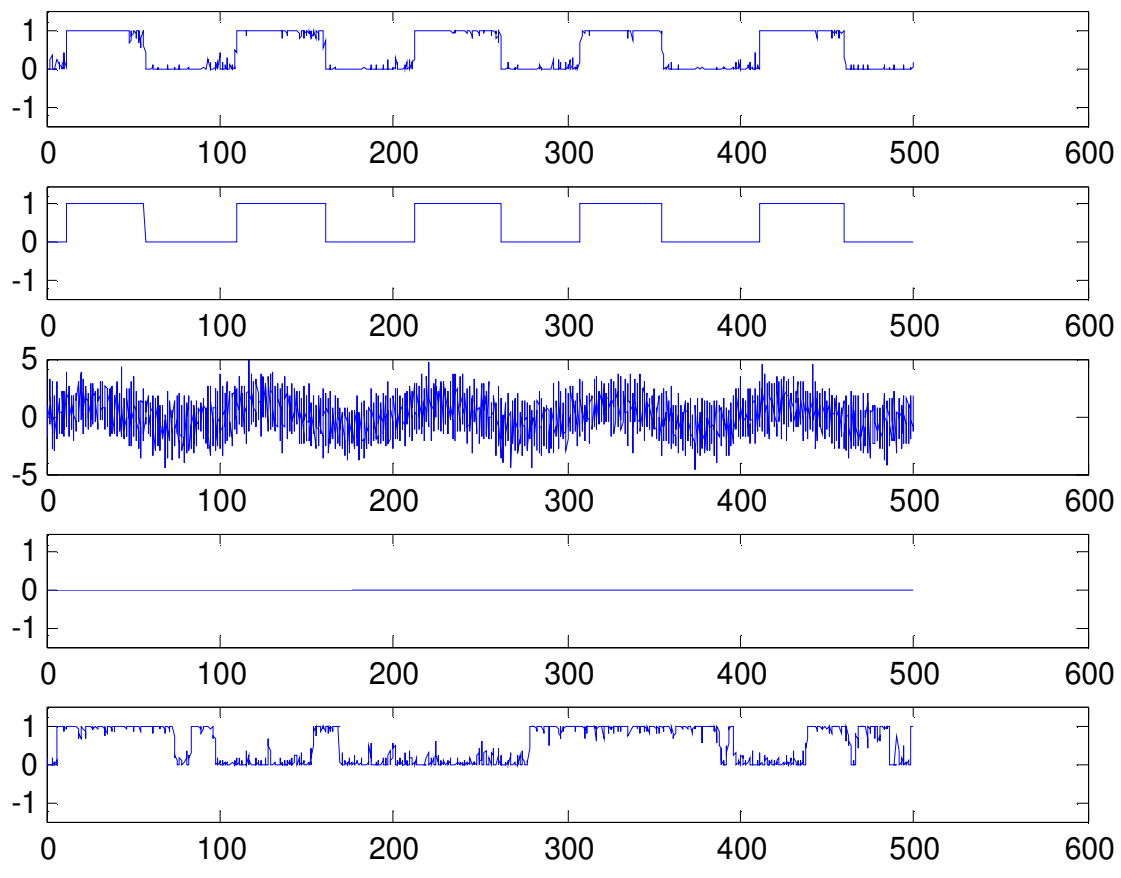

Figure 3. The simulation results for the Tsallis barrier or the roundness barrier based stochastic mechanism with $q = 4$, $f = 0.01$, $a = 1.2$, and $b = 1.0$. The explanation to the subplot 1-5 is exactly the same as that shown in the caption of Figure 2.

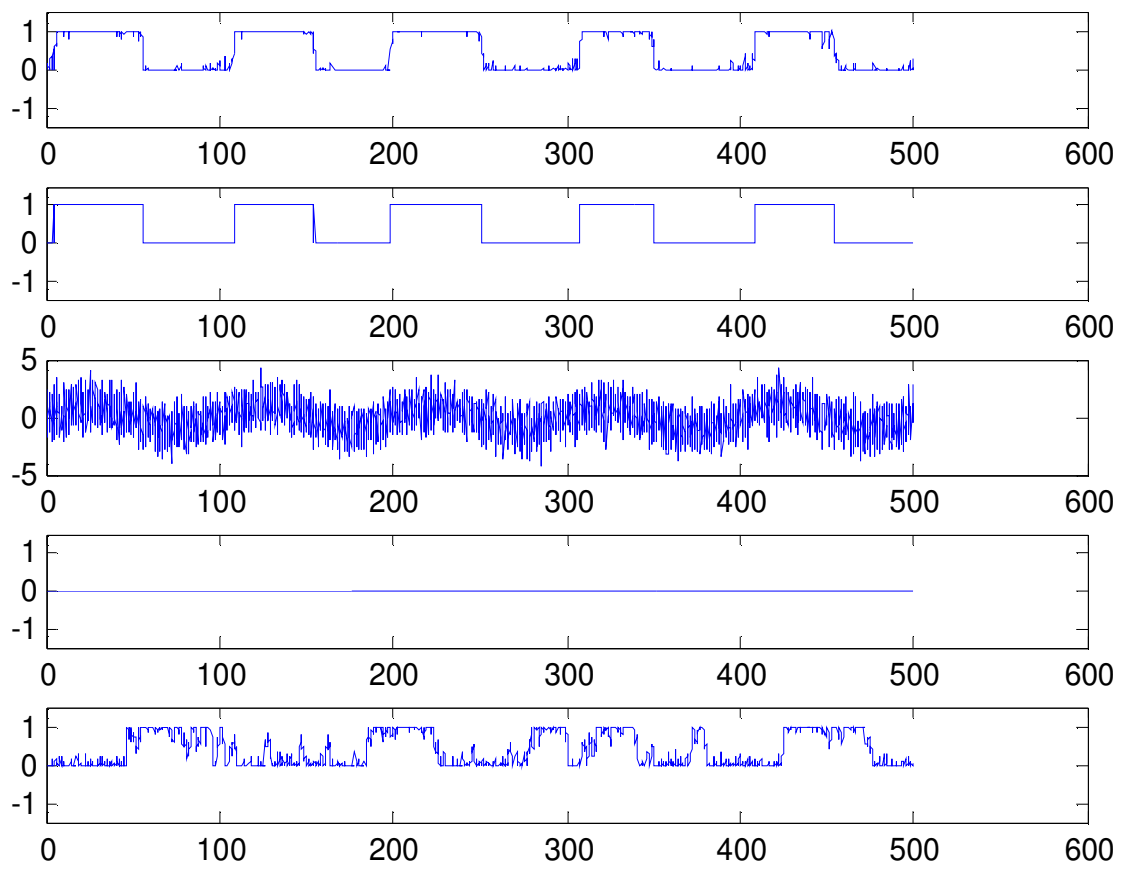

Figure 4. The simulation results for the Tsallis barrier or the roundness barrier based stochastic mechanism with $q = 10$, $f = 0.01$, $a = 1.05$, and $b = 1.0$. The explanation to the subplot 1-5 is exactly the same as that shown in the caption of Figure 2.

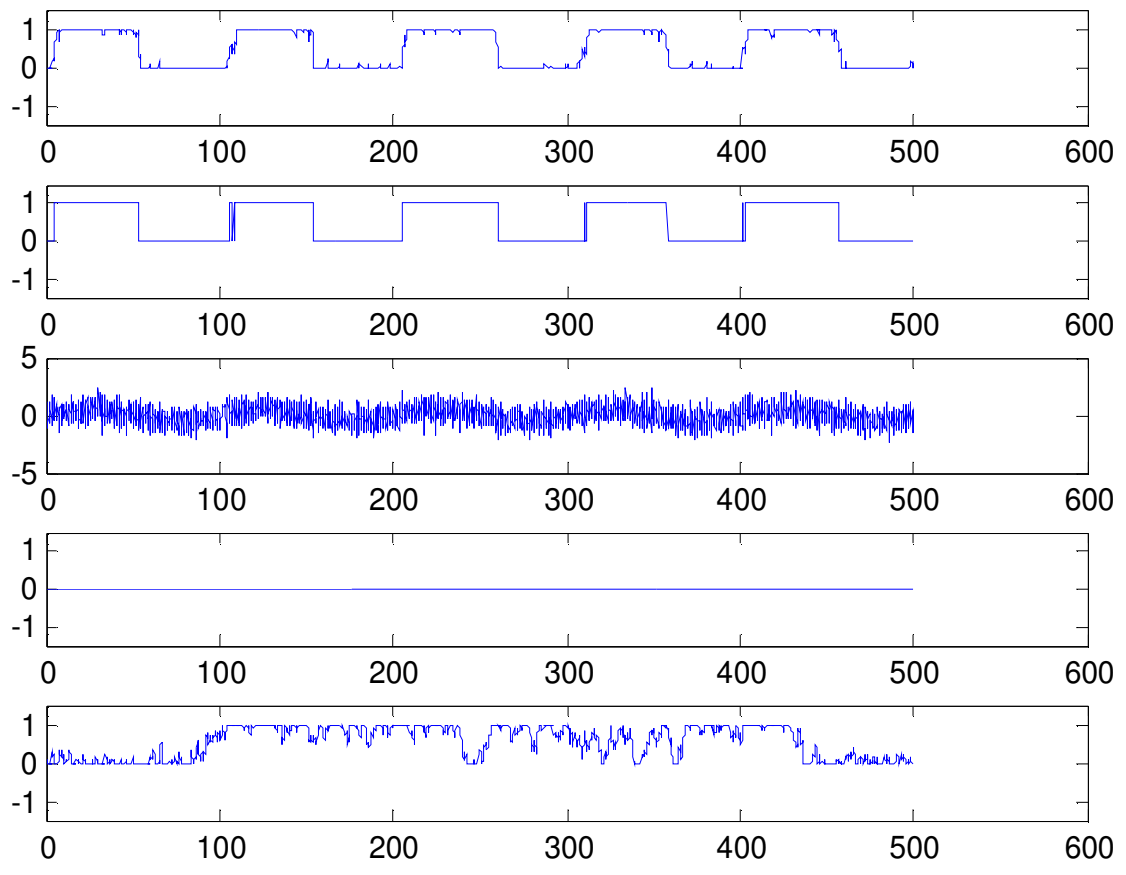

Figure 5. The simulation results for the Tsallis barrier or the roundness barrier based stochastic mechanism with $q = 50$, $f = 0.01$, $a = 0.63$, and $b = 0.5$. The explanation to the subplot 1-5 is exactly the same as that shown in the caption of Figure 2.

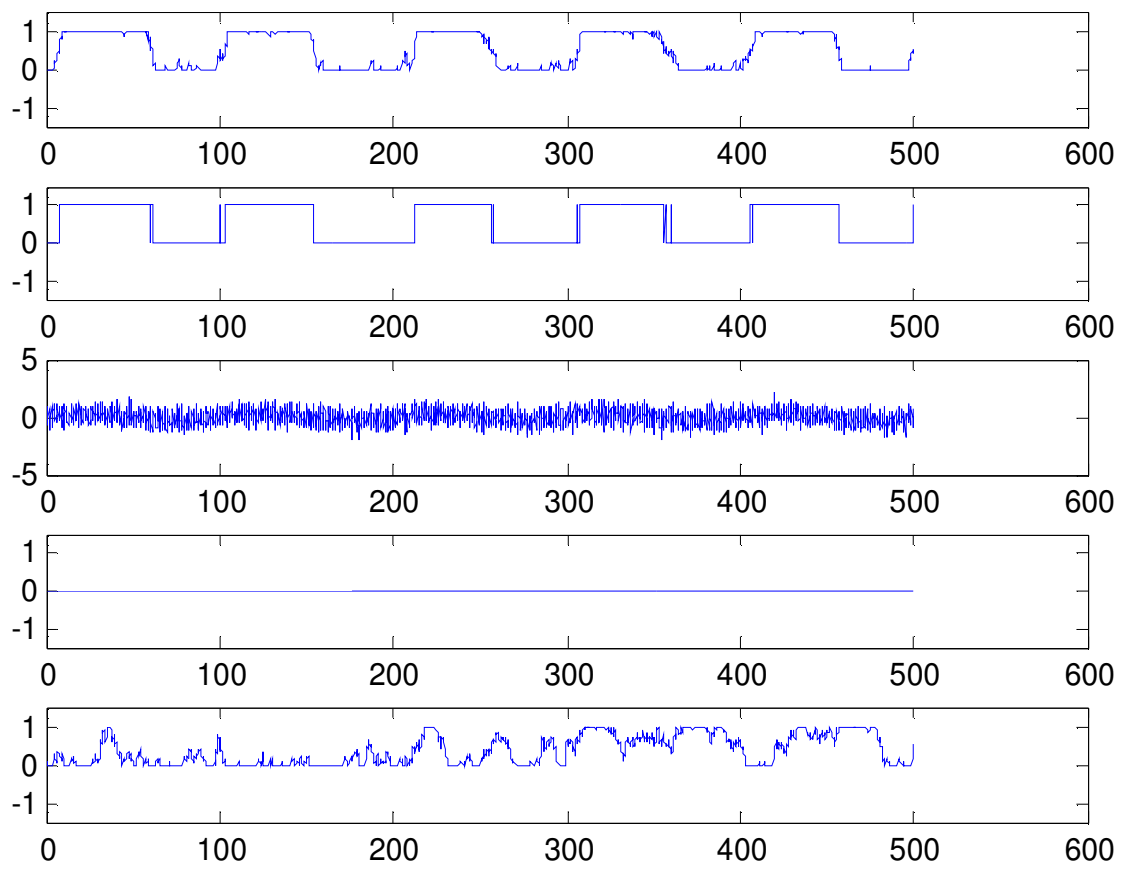

Figure 6. The simulation results for the Tsallis barrier or the roundness barrier based stochastic mechanism with $q = 1000000$, $f = 0.01$, $a = 0.52$, and $b = 0.3$. The explanation to the subplot 1-5 is exactly the same as that shown in the caption of Figure 2.

## IV. CONCLUSION

The Tsallis entropy barrier or the roundness barrier based dynamic stochastic resonance mechanisms are introduced. The simulation results for various Tsallis integer q values are presented. The systems exhibit the effects of emergence as a result of the noise-induced cooperative phenomena.

## REFERENCES

[1] R Benzi et al, “ The mechanism of stochastic resonance,” J. Phys. A: Math. Gen. **14** L453-L457, 1981.

[2] Repperger, D.W., Poo, J.A.N, “Stochastic resonance - an optimization means for manipulating uncertainty,” Intelligent Control, 2002. Proceedings of the 2002 IEEE International Symposium on, 2002**.**

[3] Roberto Fernández Galán (2006). Emergence of Brain Rhythms from Noise [PDF presentation]. Invited talk at the retreat of the CNUP (Center for Neuroscience University of Pittsburgh), Oglebay, West Virginia, USA.

[4] Cyrill B. Muratov, Eric Vanden-Eijnden, and Weinan E, "Self-induced stochastic resonance in excitable systems," Physica D 210 (2005) 227—240,2005.

[5] Tiwari, R. K., Sri Lakshmi, S. and Rao, K. N. N, " Characterization of Earthquake Dynamics in Northeastern India Regions: A Modern Nonlinear Forecasting Approach," pure & applied geophysics PAGEOPH, Vol. 161, No. 4, March 2004 , pp. 865-880(16).

[6] Thomas Wellens, Vyacheslav Shatokhin and Andreas Buchleitner, "Stochastic resonance," 2004 *Rep. Prog. Phys.* **67** 45-105.

[7] XJ Feng, JL Zhang, Q Shen, MA Tozer, “,The Ultra High Frequency Giant Stochastic Resonance-A New Effect?,” Journal – Magnetics Society of Japan, Vol 18; Number Sup//1, pages 545 1994.

[8] Xiangjun Feng, “Stochastic Resonance School of System Science,” WCFSGS, Vol.4, An Integrated Issue, No. 1-8, ISSN 1936-7260, August, 2008. (in Chinese)

[9] Xiangjun Feng, "A Unified Demonstration of the Possibility to Optimize a Few Common Uncertainties with a Simple Stochastic Resonance System," WCFSGS, Vol. 4, An Integrated Issue, No. 1-8, ISSN 1936-7260, August, 2008. (in Chinese)

[10] Xiangjun Feng, “A Public Letter to Professor Robert Vallée , the President of WOSC," WCFSGS, Vol.4, An Integrated Issue, No. 1-8, ISSN 1936-7260, August, 2008.

[11] Gingl, Z., Kiss, L. and Moss, F., "Non-dynamical stochastic resonance: theory and experiments with white and arbitrarily coloured noises", Europhys. Lett. 29, 191 (1995).

[12] Franc¸ois CHAPEAU-BLONDEAU, Julio ROJAS-VARELA, “Information-theoretic measures improved by noise in nonlinear systems,” 14th International Conference on Mathematical Theory of Networks and Systems Perpignan, France, 19—23 June 2000 ; pp. 79—82.

[13] Xiangjun. Feng, “A Theory on the determination of the roundness for generalized systems,” WCFSGS, Vol. 3, No.2, ISSN 1936-7260, Feb. 2007 (in Chinese)

[14] Xiangjun, Feng, arXiv:cond-mat.stat-mech/0705.1332 v4. May, 2007.

[15] Xiangjun Feng, “The roundness and the formula to calculate the roundness for generalized systems,” WCFSGS, Vol.4, An Integrated Issue, No. 1-8, ISSN 1936-7260, August, 2008.

[16] C. Tsallis, J. Stat. Phys. 52, (1988) 479.

[17] ET Jaynes, Phys. Rev. 106, 620 (1957).